# Nth power root topological phases in Hermitian and non-Hermitian systems

Wenyuan Deng[*], Tian Chen[*], and Xiangdong Zhang[+]

[1]Key Laboratory of advanced optoelectronic quantum architecture and measurements of Ministry of Education, Beijing Key Laboratory of Nanophotonics & Ultrafine Optoelectronic Systems, School of Physics, Beijing Institute of Technology, 100081, Beijing, China

[*]These authors contributed equally to this work. [+*]Author to whom any correspondence should be addressed. E-mail: zhangxd@bit.edu.cn; chentian@bit.edu.cn

## ABSTRACT

Constructing new topological phases is very important in both Hermitian and non-Hermitian systems because of their potential applications. Here we propose theoretically and demonstrate a general scheme experimentally to construct Nth power root (NPR) topological phases. Such a scheme is not only suitable for Hermitian systems, but also non-Hermitian systems. It is found that the robust degree of edge state in the Hermitian system becomes stronger and stronger with the increase of N. It tends to be a strongly surface localized form when N is large enough. In the non-Hermitian system, the skin effect becomes more apparent, and it approaches the ideal situation with the increase of N. This means that edge states and skin effects can be observed by taking different N. This scheme has been proved experimentally by designing circuits. Our work opens up a new way to engineer topological states according to the requirements, which is very important for developing topologically protected devices, such as topology sensing, switches, and so on.

# I. INTRODUCTION

In recent years, considerable effort has been devoted to exploring topological phases in Hermitian and non-Hermitian systems [1-3]. Topological phases can exhibit some most striking phenomena in modern physics. For example, a prominent feature of a topological phase in the Hermitian systems is the emergence of topologically protected edge states, which are robust against local perturbations and play a crucial role in the topological functionality of the underlying system [4-8]. Recent studies on topological edge states have been extended to non-Hermitian systems [9-17]. Many interesting topological phenomena associated with edge states in the non-Hermitian systems have been revealed, e.g., skin effect [18-20], which cannot be observed in Hermitian systems.

An important motivation for these studies is their potential applications [21-22]. For example, topological edge states in the Hermitian systems have potential applications in designing robust unidirectional transport devices and realizing topological quantum computation without decoherence, skin effect can be used for the topological sensing, topological switching and so on. However, how to construct more robust topological phases is the key to realizing these applications. Thus, the development of new construction methods and the construction of more robust topological phases have become a hot topic in this research field. In recent years, some novel topological phases have been proposed [23-31], the relationship between symmetry breaking and topological phases has been studied [31-38], and the effect of the interaction on the topological phase has also been discussed [39-40]. The scheme to construct square-root topological insulators from ordinary topological insulators has also been realized recently [41-46]. According to such a scheme, square-root topological states can be obtained by using a square-root procedure to the original Hamiltonian.

Motivated by the above investigations, we present a general method to construct a kind of novel topological phases, Nth power root (NPR) topological phases. In contrast to the square-root scheme, the system can still show the topological property when the NPR operation is performed for the Hamiltonian of ordinary topological insulators. Such a scheme is not only suitable for Hermitian systems, but also non-Hermitian systems. The robust degree of edge state in the Hermitian system becomes stronger and stronger with the increase of N, and it tends to be a strongly surface localized form when N is large enough. In the non-Hermitian system, the skin effect becomes more apparent, and it approaches the ideal situation with the increase of N. This means that edge states and skin effects can be designed by taking different N. This is very beneficial to developing topologically protected devices, such as topology sensing, switches, and so on.

## II. GENERAL THEORY OF NPR OPERATION

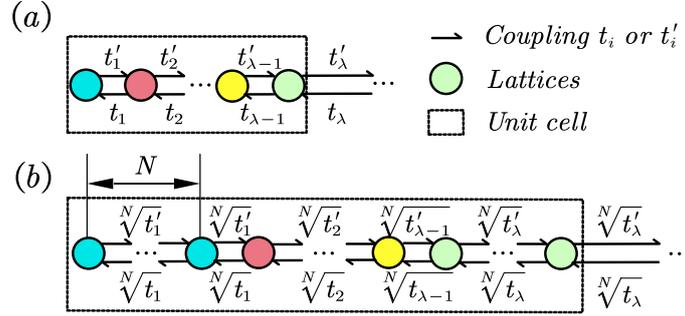

Fig. 1. The NPR operation process on a tight-binding model. The dotted box indicates a unit-cell where the coupling amplitudes and lattices are indicated by arrows and colored dots, respectively. (a) The tight-binding model. There are $\lambda$ lattices in every unit cell, and $t_\lambda$ is the coupling amplitude between cells. (b) The NPR tight-binding model. The number of lattices in every unit cell changes from $N$ to $N \cdot \lambda$, and the coupling amplitude is changed from $t$ to $\sqrt[N]{t}$.

We consider a lattice system with the nearest neighbor coupling, shown in Fig. 1(a). The colored dots and links represent lattices and coupling, respectively. The unit cell is marked by dotted lines. Here $t_i$ and $t'_i$ describe all the nearest-neighbor coupling strengths towards the left and right directions, respectively. The Hamiltonian of such a tight binding system in periodic boundary conditions (PBC) can be expressed as

$$H = \sum_{i=1}^{\lambda-1}(t_i a_i^\dagger a_{i+1} + t'_i a_{i+1}^\dagger a_i) + t'_\lambda a_1^\dagger a_\lambda e^{-ik} + t_\lambda a_\lambda^\dagger a_1 e^{ik}, \qquad (1)$$

where $a^\dagger$ and $a$ are the creation and annihilation operators. The structure of this lattice model is shown in Fig. 1(a) which has a one-to-one correspondence to Eq. (1). We can construct a new graph by introducing $N-1$ dots on each links as shown in Fig. 1(b), similar to the case for the square-root topological insulators described in Ref.[41-42]. After the lattices are expanded, the number of lattices is $N \cdot \lambda$ in every unit cell after this NPR operation, see Fig. 1(b). In such a case, we set the couplings among the lattices as $\sqrt[N]{t_i}$ and $\sqrt[N]{t'_i}$ ($i = 1,2,\ldots,\lambda$), which correspond to $t_i$ and $t'_i$ in Fig. 1(a). The Hamiltonian for this tight binding system is

$$H_r = \sum_{i=1}^{\lambda-1}\left[\sum_{j=1}^{N}\left(\sqrt[N]{t_i}a_\alpha^\dagger a_{\alpha+1} + \sqrt[N]{t'_i}a_{\alpha+1}^\dagger a_\alpha\right) + \sum_{j=1}^{N-1}\left(\sqrt[N]{t_\lambda}a_\beta^\dagger a_{\beta+1} + \sqrt[N]{t'_\lambda}a_{\beta+1}^\dagger a_\beta\right)\right]$$

$$+ \sqrt[N]{t_\lambda}a_1^\dagger a_{N\lambda}e^{-ik} + \sqrt[N]{t'_\lambda}a_{N\lambda}^\dagger a_1 e^{ik}, \qquad (2)$$

where $= N(i-1) + j$, $\beta = N(\lambda-1) + j$. The $H_r$ can be regarded as the NPR of the original Hamiltonian $H$. We can prove that the $H_r$ and $H$ have the same topological properties, the detailed demonstration is given in Appendix A.

Eqs. (1) and (2) describe the case of general tight-binding systems. As $t_i = t'_i$, it corresponds to the Hermitian case, $t_i \neq t'_i$ to the non-Hermitian case. In addition, when $\lambda = 2$, Eqs. (1) and (2) transform to the Hamiltonians of the SSH model. In the following, we take the SSH model as an example to discuss the topological properties of $H_r$ for Hermitian and non-Hermitian cases in detail.

### III. NPR HERMITIAN SSH MODEL

In the Hermitian case ($t_i = t'_i$), the NPR Hermitian SSH model is given by

$$H_{r,\lambda=2} = \sum_{i=1}^{N} \sqrt[N]{t_1} a_i^\dagger a_{i+1} + \sum_{i=1}^{N-1} \sqrt[N]{t_2} a_{N+i}^\dagger a_{N+i+1} + \sqrt[N]{t_2} a_1^\dagger a_{2N} e^{-ik} + h.c., \quad (3)$$

which corresponds to Eq. (2) with $\lambda = 2$. The eigenvalues of Eq. (3) can be calculated by matrix diagonalization, from which we can get the energy spectrum for arbitrary $N$. As an example, in Fig. 2(a), we plot the energy spectrum of the SSH model for the 7$^{th}$ root (N=7), where two energy bands (marked in red) are degenerate in the region $t_1 < t_2$. For comparison, the corresponding energy spectrum for the original SSH model is shown in Fig. 2(b). It is found that the transition points of both cases appear at $t_1 = t_2$. The degenerate eigenvalues appear in the energy gap, although the regions for the finite energy gap are different for the two cases. This is because the NPR operation increases the order of the matrix in solving eigenvalues [28], which leads to the shift of the energy spectrum. But it does not affect that original and NPR models have the same physical properties. This can be seen more clearly in Fig. 2(c). Figure 2(c) shows the density of states (DOS) as a function of cell index $m$. The red line and green line correspond to the case of the original SSH model and the 7th root SSH model, respectively. It is seen clearly that the states for two cases are localized at the chain boundary. The isolated energies are indicated by an ellipse in the inset of Fig. 2(c), which means the state is topologically protected. The topologically protected properties of the state in the original model have been discussed in Ref. [5]. In fact, we can also calculate the winding number [47-49] of the NPR SSH model to determine the topological region ($t_1 \leqslant t_2$). The detailed calculation process can be found in S1 of Supplementary Material. The difference between them is that the topologically protected state for the NPR model is more localized than that for the original model marked in red. To quantitatively characterize the localization properties of these boundary states, the inverse participation ratio (IPR) [38] is introduced as

$$\eta = \sum_r |\psi_\alpha(r)|^4 / (\sum_r |\psi_\alpha(r)|^2)^2 \quad , \quad (4)$$

where $\psi_\alpha(r)$ represents all localized eigenstates, and $r$ contains the contribution from all lattices in $m = 1$. The closer IPR is to 1, the higher the degree of localization of the system. The red line in Fig. 2(d) describes the calculated results of IPR as a function of N. It is found that the $\eta$ increases monotonically with the increase of $N$ in the case of small $N$, and then converges to 1 gradually. As a result, the topological enhancement effect and the rapid convergence of localization degree in the NPR system can be widely used in the design of quantum devices, such as topology switches, robust transport and so on.

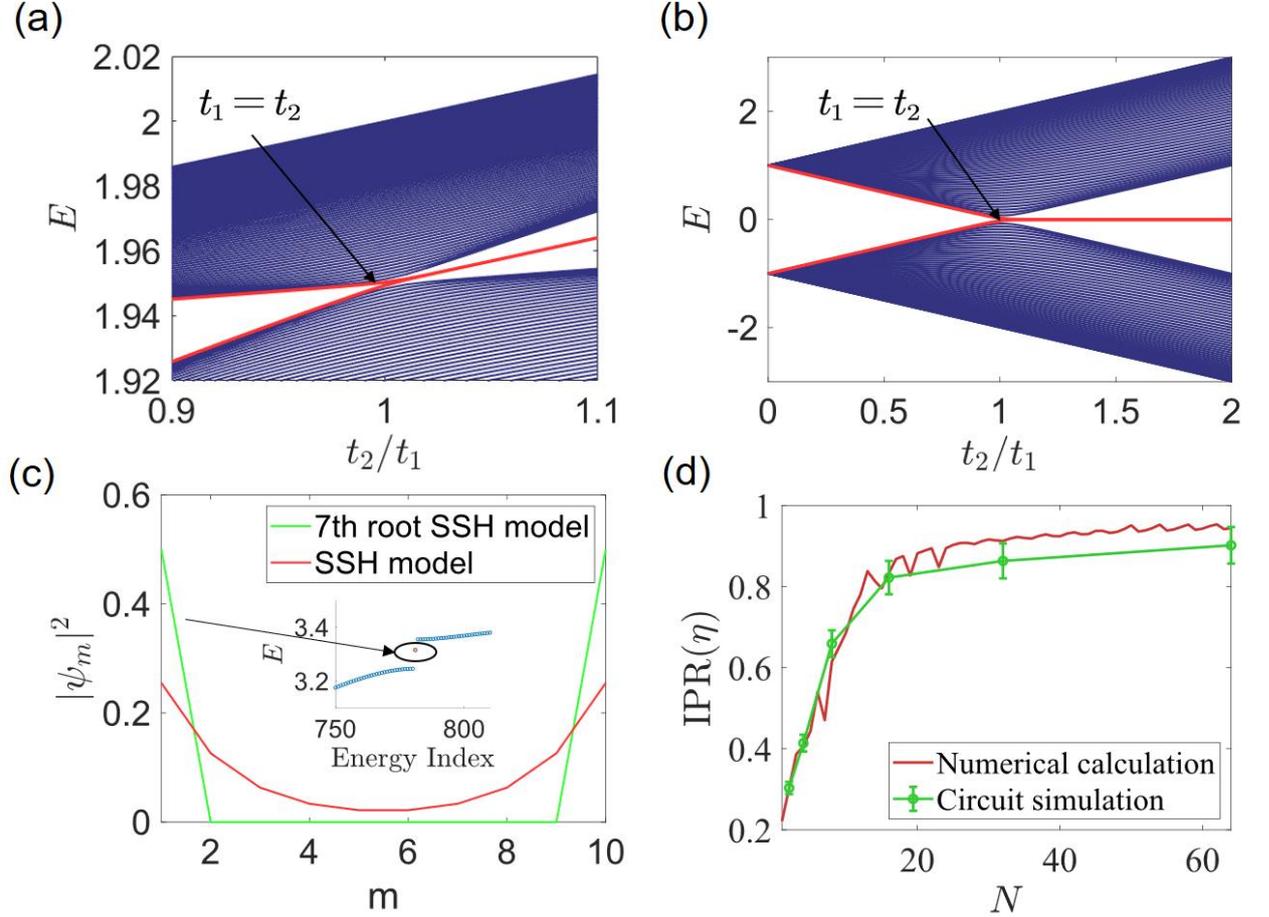

Fig. 2. Bulk–boundary correspondence (BBC) in Hermitian NPR model. The open boundary condition (OBC) is chosen. (a) The numerical spectrum of the 7$^{th}$ root SSH model with length L = 40 (unit cell). $t_1 = 1, t_2$ varies in $[0,2]$. $|E|$ as functions of $t_1$. The isolated energies corresponding to the localized eigenstates are finite, which are shown in red (twofold), and the transition point is indicated by an arrow. (b) The same as (a) except that the model is replaced by the original SSH model. The transition point is still at $t_1 = t_2$, but the isolated energies for the localized eigenstates are zero (shown in red). (c) Boundary states with length L = 10 (unit cell). $t_1 = 33, t_2 = 47$. $|\psi_m|^2$ is the density of states (DOS) of corresponding unit cell $m$, and $m$ is the index of each unit cell on the chain. The boundary states of the SSH model and the 7$^{th}$ root SSH model are shown in red and green, respectively. The isolated energies are indicated by an ellipse in the inset of Fig. 2(c). (d) The relationship between IPR and the times (N) with length L = 10 (unit cell). $t_1 = 33, t_2 = 47$. The theoretical calculation result is shown in red, and $N$ varies in $[1,64]$. The circuit simulation result is in green and $N = 2,4,8,16,32,64$.

In the simulation, the set error of components is 10%. The results of the two calculation methods can match well within the simulation error range.

The above phenomenon can also be observed experimentally by designing circuit networks [50-62]. Recent research has shown that the circuit Laplacian has the same expression as the Hamiltonian of the lattice system when the circuit is appropriately designed. The theory of circuit design corresponding to Eq. (3) is given in S2 (A) of Supplementary Material. The designed and fabricated circuit networks are shown in Fig. 3(a), containing 167 solderable nodes. Among these nodes, we can set different components to simulate the couplings in model systems according to the admittance matrix. For example, inset (i) in Fig. 3(a) shows the unit cell of the designed circuit network for the case with N=2, which consists of four capacitances and four inductances. For the case with N=4, it contains eight capacitances and eight inductances, as shown in inset (ii) of Fig. 3(a). Similarly, for the case with N=8, the unit cell consists of sixteen capacitances and sixteen inductances as shown in inset (iii) of Fig. 3(a). In inset (iii), the different capacitances $C_1$ and $C_2$ are marked by yellow and red colors, respectively, and inductances are not marked. The experimental parameters are also shown in Fig. 3(a), which are the same for the three cases mentioned above. As a result, we can get three experiment results by controlling the number of components on only one circuit board.

Before doing experiments, we first perform numerical simulations (with the software LTspice) of IPR and state distributions on the designed circuit networks with various sizes. The simulation result of IPR with different N is shown in Fig. 2(d) (marked in green). It is shown clearly that the circuit simulation results are basically consistent with the theoretical results, some differences come from the loss of circuit components. The simulation results of state distributions for the systems with N=2, 4 and 8 are shown in Fig. 3(b) (green). The red lines are corresponding practical measurement results. It is seen clearly that both the experimental and simulation results exhibit a strong localization effect around the boundary. For example, the experimental measurement results of the IPR are 0.3019, 0.4011 and 0.6622 for N=2, 4 and 8, respectively, and the corresponding simulation results are 0.3033, 0.4140 and 0.6593. Within the error range, the consistencies between experimental and simulation results are very well.

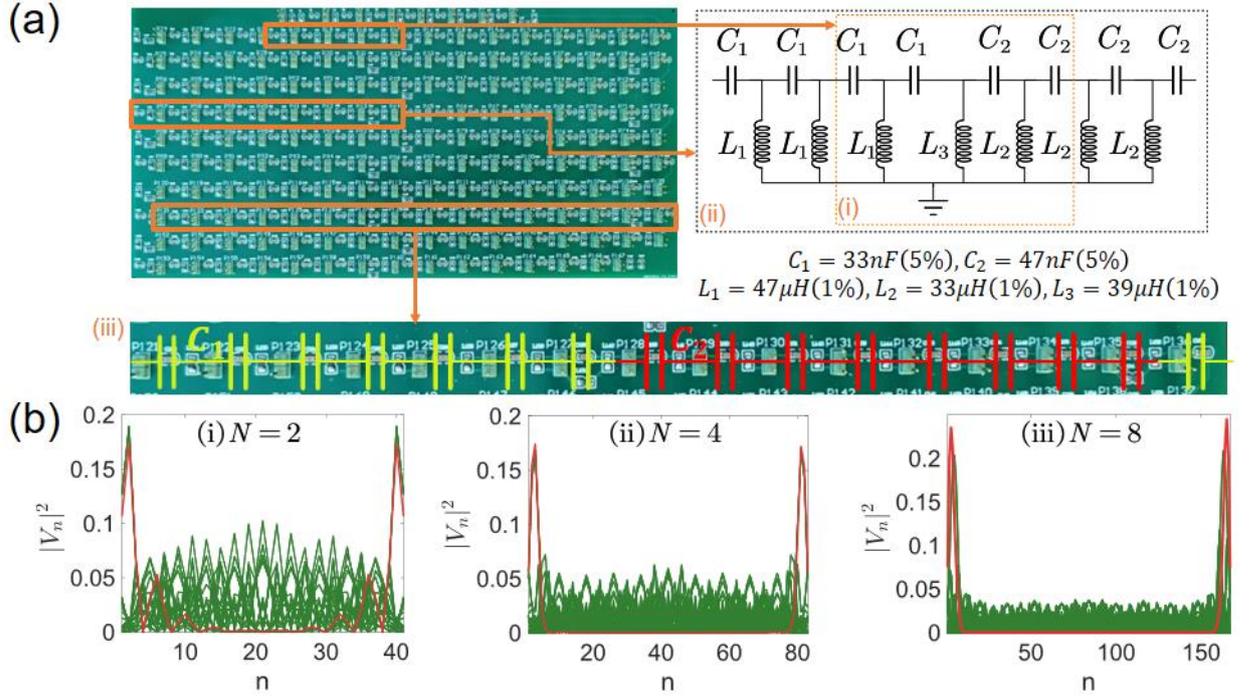

Fig. 3 The photograph of the circuit board and experimental results for the NPR SSH model. (a) Circuit board cutout of the Hermitian circuit experiment. The circuit board contains 167 solderable nodes. The insets (i) and (ii) are the circuit diagrams for $N = 2$ and $N = 4$, respectively. The inset (iii) is the cutout of one unit-cell for $N = 8$. (b) The boundary states in circuit experiments occur at corresponding frequencies $f \approx 90.365 \text{kHz}$. $|V_n|^2$ is the voltage density of node index n in the circuit system. The circuit simulation results include errors considered are shown in green, and experiment results are in red. As seen in (i) to (iii), the boundary state is more localized, where the $IPR_{(i)} = 0.3019$, $IPR_{(ii)} = 0.4011$, $IPR_{(iii)} = 0.6622$.

## IV. NPR NON-HERMITIAN SSH MODEL

Similar to Hermitian cases, the NPR operation can also keep the topological phase in non-Hermitian cases ($t_i \neq t_i'$). From Eq. (2), Hamiltonian of the non-Hermitian system is expressed as

$$H'_{r,\lambda=2} = \sum_{i=1}^{N-1} \sqrt[N]{t_1 + \gamma}\, a_i^\dagger a_{i+1} + \sum_{i=1}^{N-1} \sqrt[N]{t_1 - \gamma}\, a_{i+1}^\dagger a_i + \sum_{i=1}^{N-1}\left(\sqrt[N]{t_2}\, a_{N+i}^\dagger a_{N+i+1} + \sqrt[N]{t_2}\, a_1^\dagger a_{2N} e^{-ik} + h.c.\right), \quad (5)$$

where $t_1 + \gamma$ and $t_1 - \gamma$ represent the couplings in the non-Hermitian SSH model corresponding to $t_i \neq t_i'$. According to Eq. (6), we can calculate the energy spectrum of arbitrary $N$. As an example, in Fig. 4(a), we plot the spectrum of the SSH model with the OBC for $N = 5$, where the isolated energy is marked by a red line and the transition points are indicated by arrows [$t_1 \approx \pm\sqrt{t_2^2 + \gamma^2}$]. The corresponding energy spectrum for the original SSH model is shown in Fig. 4(b). Comparing Fig. 4(a) with Fig. 4(b), it is found that the transition points of both cases appear at $t_1 \approx \pm\sqrt{t_2^2 + \gamma^2}$. As the corresponding DOSs of $N = 2$ and 5 are shown in the and $N = 2, 5$. Localized states at the right boundary are not shown. Those of $N = 5$ marked in green are shaper than the states of $N = 2$ marked in blue. The inset is the spectrum of $N = 2$. It shows the same topological property as (a) and (b). (d) The IPR as functions of N with L = 10 (unit cell). The numerical results are shown in red and $N$ varies in [1,32]. The circuit simulation results are in green and $N = 2,4,8,16,32$. In the simulation, the set error of components is 10%. The results of the two calculation methods can match well within the simulation error range.

Fig. 4(c), which are as a function of cell index m. The blue dotted line and green line correspond to the case of the square SSH model and the 5th root SSH model, respectively. It is seen clearly that the skin effect is shown in both cases. Corresponding to the DOS in the case of $N = 2$, the spectrum is shown in the inset of Fig. 4(c). To quantitatively characterize the localization properties of the skin effect in NPR models, we also calculate the IPR of the NPR non-Hermitian SSH model. The blue line in Fig. 4(d) describes the computed results of IPR as a function of N. It is found that the η increases monotonically with the increase of N when N is small, and then converges to 1 gradually. Similar to the Hermitian cases, the NPR operation can enhance the localization of the topologically protected state. As a result, the topological enhancement effect and the rapid convergence of localization degrees in the NPR system can also be widely used in the design of quantum devices.

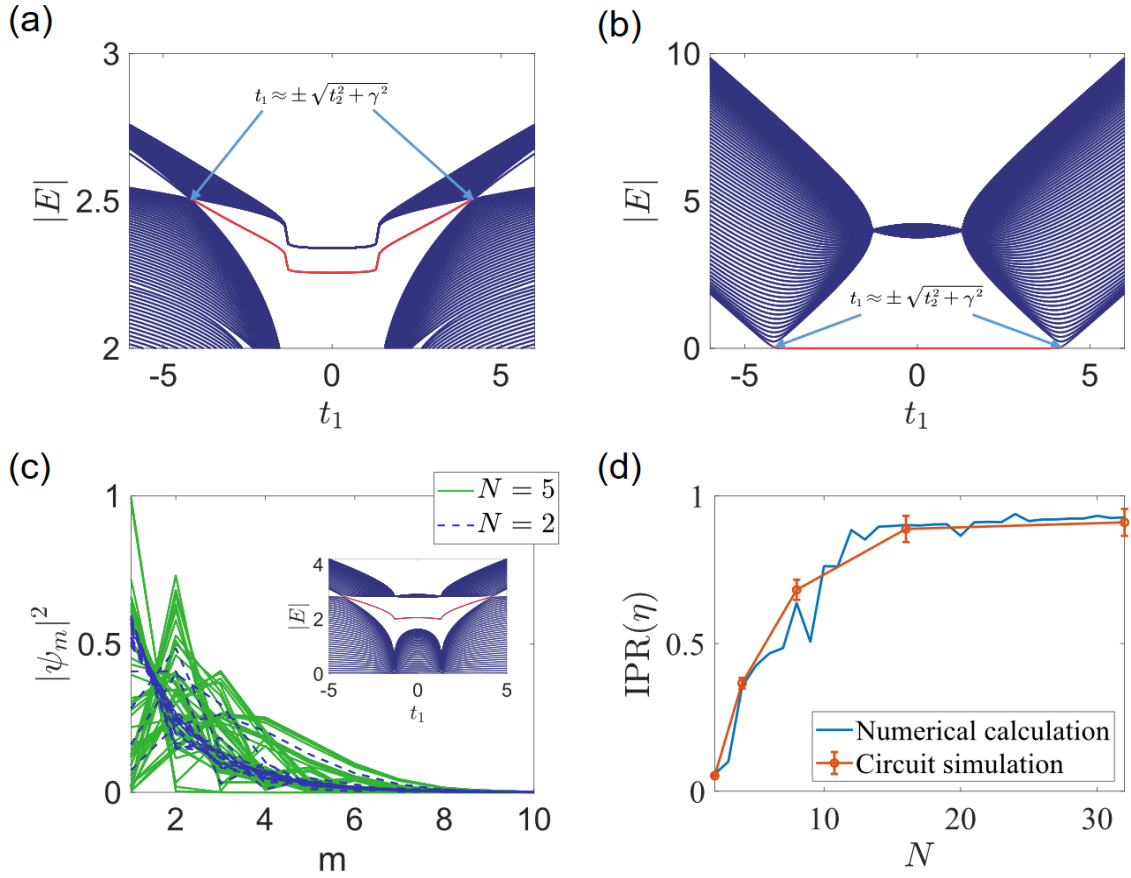

Fig. 4. Bulk–boundary correspondence (BBC) in the NPR non-Hermitian model. The OBC is chosen. (a) The numerical spectra of the non-Hermitian 5th root SSH model with the length L = 40 (unit cell). $t_2 = 4, \gamma = 4/3$ and $t_1$ varies in [-5, 5]. The topological mode line marked in red is between the transition points $[t_1 \approx \pm\sqrt{t_2^2 + \gamma^2}]$ indicated by arrays. (b) The numerical spectra of the original non-Hermitian SSH model with the length L = 40 (unit cell). $t_2 = 4, \gamma = 4/3$ and $t_1$ varies in [-5, 5]. Same as (a), the topological mode is marked in red, and the transition points $[t_1 \approx \pm\sqrt{t_2^2 + \gamma^2}]$ are indicated by arrays. (c) The skin effect of the NPR non-Hermitian SSH model with L = 10 (unit cell). $t_2 = 4, \gamma = 4/3$

According to the latest research [47-48, 63-67], the usual bulk-boundary correspondence breaks down in the non-Hermitian systems. So, we have to study further the properties of NPR models in the momentum space and compare the properties before and after the NPR operation. In this process, it is found that the spectra of NPR models in the momentum space drift from the original position. So a correction is required to eliminate the unwanted drift. The more information about unusual BBC and the correction of NPR operation for momentum space, see S3 of Supplementary Material.

Similar to the Hermitian cases, the skin effect also can be observed experimentally by designing the circuit networks. But different from Hermitian cases, there are unequal couplings in non- Hermitian systems. So, we need to use a particular component to express the non-Hermitian coupling, which is the negative impedance converter with a current inversion circuit (INIC) [68]. More information about the design of INIC is shown in supplementary material S2 (B). The circuit board cutout is demonstrated in Fig. 5(a). It is for N=4, and the circuit design of N=2 can be found in S2 (B) of supplementary material. We mark a unit cell in Fig. 5(a) with an orange square and enlarge it as shown in Fig. 5(b). In Fig. 5(b), the INIC is marked in red which simulates the unequal coupling in the non-Hermitian system. The capacitance marked in purple simulates the coupling $\sqrt[N]{t_1}$ in Eq. (6). And the grounding elements marked in yellow and blue are grounding elements in the Hermitian part and non-Hermitian part, respectively. Based on the designed circuit network, we first perform numerical simulations of IPR and DOSs on the designed circuit networks with different sizes. The simulation result of IPR with various N is shown in Fig. 4(d) (marked in orange). It is shown clearly that the circuit simulation results are basically consistent with the theoretical results, some differences come from the loss of circuit components.

The difference between PBC and OBC is just coupling between the first and last nodes. As a result, we can measure the circuit properties on the same board, the detailed measuring method can be found in S2 (A) of supplementary material. As the measuring results for N=2 are shown in Fig. 5(c-1), the apparent skin effect has been shown. And from the inset of Fig. 5(c-1), we can see the difference between PBC and OBC spectra that the PBC spectrum traces out closed loops in the complex admittance plane while the OBC spectrum is in a line. The measuring spectrum corresponds to the skin effect on the circuit and shows the unusual BBC in the non-Hermitian system. And on the circuit platform, the principal notion of skin effect would also suggest a voltage profile. So, we set the current feed at the last nodes of the circuit, and then we can find that the predominant weight of current is not at the last but at the beginning of the circuit, as shown in Fig. 5(c-2). The same properties also can be seen in the quartic root system, $N = 4$. As shown in Fig. 5(d) ($N = 4$), the skin effect and the unusual BBC are also kept in this system. Moreover, the skin effect of $N = 4$ is shaper than that of $N = 2$, with the IPRs of both two systems, $\eta_2 = 0.1241$ and $\eta_4 = 0.4121$. Compared with the simulation results in Fig. 4(d), these two values of IPR are 0.0577 and 0.3656 in the experiment, which displays a good agreement.

From the experiments in non-Hermitian cases, we can ensure the conjecture in theoretical derivation and find that the enhancement of skin effect with times $N$ can be realized in a classical circuit. The skin effect phenomenon shown in Fig. 5(c-2) and Fig. 5(d-2) has a good application prospect in the sensor research and development.

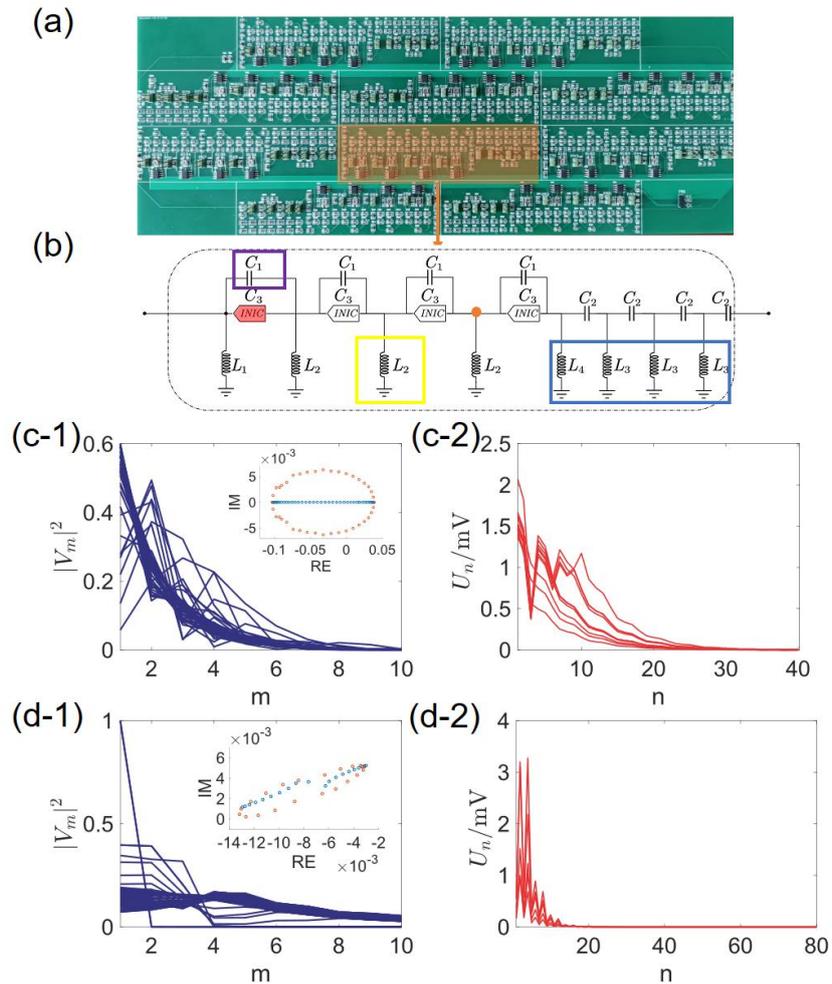

Fig.5. The experiment design and result of the NPR non-Hermitian SSH model. The OBC is chosen. (a) The circuit board of $N = 4$ with L = 10 (unit cell). A unit cell is marked by an orange square corresponding to the circuit diagram in Fig. 5(b). (b) Simplified circuit diagram of $N = 4$. One of the measuring nodes is marked by the orange dot. An INIC is marked in red. The capacitance is marked by the purple color. The capacitance $C_3$ in the INIC circuit is for adjusting the forward and reverse admittance. And the grounding elements marked in yellow and blue are grounding elements in the Hermitian part and non-Hermitian part, respectively. The parameters of main components in experiment are as followed, $C_1 = 220nF, C_2 = 223.3nF, C_3 = 19.3875nF, L_1 = 100\mu H, L_2 = 98.4\mu H, L_3 = 103.3\mu H, L_4 = 95\mu H$ and $R_1 = 5.37\Omega, R_2 = 21.62\Omega, R_3 = 21.33\Omega$. (c) Experimental results for the skin effect in the square root model occur at the frequency f≈30kHz. (c-1) $|V_m|^2$ is the voltage density of cell index m in the circuit system. (c-1) $U_n$ is the voltage of the corresponding measuring node index $n$. The spectrum in the complex space is shown in the inset of (c-1), where the spectrum in the PBC is marked in orange, and the spectrum in the OBC is marked in blue. (d) Experimental results for the skin effect in the quartic root model occur at the frequency f≈24kHz. The IPRs calculated respectively in (c) and (d) are $IPR_{(i)} = 0.1241$ and $IPR_{(ii)} = 0.4121$.

## V. SUMMARY

We have presented a general scheme to construct NPR topological phases. Such a scheme is not only suitable for Hermitian systems, but also for non-Hermitian systems. We have demonstrated that the topological properties of the NPR model are consistent with the topological properties of the original model. We have also found that the robust degree of edge state in Hermitian systems becomes stronger and stronger with the increase of N, and it tends to be a strongly surface localized form when N is large enough. In non-Hermitian systems, the skin effect becomes more apparent, it approaches the ideal situation with the increase of N. This means that edge states and skin effect can be designed by taking different N. Furthermore, we have constructed the Hermitian and non-Hermitian circuit networks for the SSH model to demonstrate the theoretical scheme experimentally It is expected that the operation method we have provided can be implemented in more topological materials, which has potential application value in the design of topologically protected devices. The detailed phenomenon is given in Appendix B.

Note added. —We became aware of a recent related theoretical work [69] that focuses on the qth-root non-Hermitian Floquet topological insulators.

## ACKNOWLEDGMENTS

This work was supported by the National Key R & D Program of China under Grant No. 2017YFA0303800, the National Natural Science Foundation of China (91850205).

## APPENDIX A: PROOF OF PROPERTY OF NPR OPERATION

We have constructed the NPR Hamiltonian $H_r$ on the 1D tight-binding model through the subdivided graph. Then we can decompose it as $H'$:

$$H' = U_0 H_r U_0^{-1} = \begin{pmatrix} O_\lambda & H_\alpha & 0 & & & H'_\alpha \\ H_\beta & O_\lambda & H_\alpha & 0 & & \\ 0 & H_\beta & \ddots & \ddots & \ddots & \\ & 0 & \ddots & \ddots & H_\alpha & 0 \\ & & \ddots & H_\beta & O_\lambda & H_\alpha \\ H'_\beta & & & 0 & H_\beta & O_\lambda \end{pmatrix}_N, \tag{A1}$$

which is a block matrix with a block size of $\lambda$. $U_0 = \sum_{i=1}^{\lambda} \sum_{j=1}^{N} a^\dagger_{\lambda(j-1)+i} a_{N(i-1)+j}$. $O_\lambda$ is a $\lambda$-dimensional zero matrix, and the others are as follows,

$$H_\alpha = \begin{pmatrix} \sqrt[N]{t_1} & 0 & 0 & 0 \\ 0 & \sqrt[N]{t_2} & \ddots & 0 \\ 0 & \ddots & \ddots & 0 \\ 0 & 0 & 0 & \sqrt[N]{t_\lambda} \end{pmatrix}, H'_\alpha = \begin{pmatrix} 0 & 0 & 0 & 0 & \sqrt[N]{t'_\lambda} e^{-ik} \\ \sqrt[N]{t'_1} & 0 & 0 & 0 & 0 \\ 0 & \sqrt[N]{t'_2} & \ddots & \ddots & 0 \\ 0 & \ddots & \ddots & 0 & 0 \\ 0 & 0 & 0 & \sqrt[N]{t'_{\lambda-1}} & 0 \end{pmatrix},$$

$$H_\beta = \begin{pmatrix} \sqrt[N]{t'_1} & 0 & 0 & 0 \\ 0 & \sqrt[N]{t'_2} & \ddots & 0 \\ 0 & \ddots & \ddots & 0 \\ 0 & 0 & 0 & \sqrt[N]{t'_\lambda} \end{pmatrix}, H'_\beta = \begin{pmatrix} 0 & \sqrt[N]{t_1} & 0 & 0 & 0 \\ 0 & 0 & \sqrt[N]{t_2} & 0 & 0 \\ 0 & 0 & \ddots & \ddots & 0 \\ 0 & \ddots & \ddots & 0 & \sqrt[N]{t_{\lambda-1}} \\ \sqrt[N]{t_\lambda} e^{ik} & 0 & 0 & 0 & 0 \end{pmatrix}. \tag{A2}$$

After blocking, we use the Nth power of $H'$ to track the original system,

$$(H')^N = H_d + H_{enh}, \tag{A3}$$

in which, $H_d = H_d^1 \oplus H_d^2 \cdots \oplus H_d^N$ is a diagonal blocking Hamiltonian, $H_d^i$ is a $\lambda$-dimensional term of $H_d$ and $H_{enh}$ is the enhanced term. As mentioned in the text, there is an enhancement phenomenon in the topological phase within a certain limit. For a detailed description of the enhancement term, you can refer to the numerical calculation in the text and the model analysis in S4 of Supplementary Material.

The expression of $H_d^i$ is as follows,

$$H_d^i = (H_\alpha)^{N-i} H'_\beta (H_\alpha)^{i-1} + (H_\beta)^{i-1} H'_\alpha (H_\beta)^{N-i}, \tag{A4}$$

when $i = 1$, the first block is $H_d^1 = (H_\alpha)^{N-1} H_\beta' + H_\alpha' (H_\beta)^{N-1}$, which is identical to the original Hamiltonian up to an additive constant. At the same time, we can give its determinant as,

$$|H_d^1| = e^{ik} \cdot \prod_{j=1}^{\lambda} t_j + e^{-ik} \cdot \prod_{j=1}^{\lambda} t_j' + C \tag{A5}$$

$$|H_d^1| = |(H_\alpha)^{N-1} H_\beta'| + |H_\alpha' (H_\beta)^{N-1}| + C_{t_1, t_2, \cdots, t_\lambda, t_1', t_2', \cdots, t_\lambda'} \tag{A6}$$

$$|(H_\alpha)^{N-1} H_\beta'| = e^{ik} \cdot \prod_{j=1}^{\lambda} t_j \tag{A7}$$

$$|H_\alpha' (H_\beta)^{N-1}| = e^{-ik} \cdot \prod_{j=1}^{\lambda} t_j' \tag{A8}$$

in which, $C$ is a constant independent of wave vector k and has different forms in the parity of $\lambda$,

$$C = \begin{cases} \prod_{j=1}^{\lambda/2} t_{2j-1} t_{2j-1}' + \prod_{j=1}^{\lambda/2} t_{2j} t_{2j}', & \lambda \text{ is even} \\ 0, & \lambda \text{ is odd} \end{cases}. \tag{A9}$$

We can extend the above derivation to $H_d^i$, the left of $H_d^i$ is as follows,

$$(H_\alpha)^{N-i} H_\beta' (H_\alpha)^{i-1} = (H_\alpha)^{1-i} [(H_\alpha)^{N-1} H_\beta'] (H_\alpha)^{i-1}. \tag{A10}$$

It obtains that the left of $H_d^i$ is similar to $(H_\alpha)^{N-1} H_\beta'$, they have equal determinant. Same as the right,

$$(H_\beta)^{i-1} H_\alpha' (H_\beta)^{N-i} = (H_\beta)^{i-1} [H_\alpha' (H_\beta)^{N-1}] (H_\beta)^{1-i}. \tag{A11}$$

Then we can give $H_d^i$'s determinant like $H_d^1$ as,

$$|H_d^i| = |(H_\alpha)^{N-i} H_\beta' (H_\alpha)^{i-1}| + |(H_\beta)^{i-1} H_\alpha' (H_\beta)^{N-i}| + C'_{t_1, t_2, \cdots, t_\lambda, t_1', t_2', \cdots, t_\lambda'}$$

$$= |(H_\alpha)^{N-1} H_\beta'| + |H_\alpha' (H_\beta)^{N-1}| + C'_{t_1, t_2, \cdots, t_\lambda, t_1', t_2', \cdots, t_\lambda'}, \tag{A12}$$

where,

$$C' = \begin{cases} \Delta \cdot \prod_{j=1}^{\frac{\lambda}{2}} t_{2j-1} t_{2j-1}' + \Delta^{-1} \cdot \prod_{j=1}^{\frac{\lambda}{2}} t_{2j} t_{2j}', & \lambda \text{ is even} \\ 0, & \lambda \text{ is odd} \end{cases}, \tag{A13}$$

and $\Delta = \prod_{j=1}^{\lambda/2} \left( \frac{t_{2j} t_{2j}'}{t_{2j-1} t_{2j-1}'} \right)^{\frac{i-1}{N}}$. Comparing Eq. (A12) and (A6): when $\lambda$ is odd, $H_d^i$ has the same determinant. Also, because they have the same coupling structure and trace $Tr(H_d^i) = 0$, $H_d^i$ has the same eigenvalues. $H_d$ is a diagonal blocking matrix, and we can prove that $H_d$ has the same eigenvalues as $H_d^1$:

$$|H - x \cdot I| = \begin{vmatrix} H_1 - x \cdot I_1 & & & \\ & H_2 - x \cdot I_2 & & \\ & & \ddots & \\ & & & H_N - x \cdot I_N \end{vmatrix}$$

$$= |H_1 - x \cdot I_1||H_2 - x \cdot I_2| \cdots |H_N - x \cdot I_N| = \prod_{i=1}^{N} |H_i - x \cdot I_i|, \tag{A14}$$

when $x$ is the eigenvalue for the block $H_i$, so is $H$. When $\lambda$ is even, to prove the topological phase has no change, we need to give the matrix form of $H_d^i$ as follows,

$$H_d^i = \begin{pmatrix} 0 & \gamma_1 t_1 & & & & \gamma_\lambda' t_\lambda' e^{-ik} \\ \gamma_1' t_1' & 0 & \gamma_2 t_2 & & & \\ & \gamma_2' t_2' & 0 & \ddots & & \\ & & \ddots & \ddots & & \gamma_{\lambda-1} t_{\lambda-1} \\ \gamma_\lambda t_\lambda e^{ik} & & & & \gamma_{\lambda-1}' t_{\lambda-1}' & 0 \end{pmatrix}, \tag{A15}$$

where $\gamma_j = \left(\frac{t_{j+1}}{t_j}\right)^{\frac{j-1}{N}}$ ($j < \lambda$; when $j = \lambda$, $\gamma_\lambda = \frac{t_1}{t_\lambda}$), $\gamma'(t \to t')$. So $H_d^i$ has the same form as $H_d^1$ at point $t_i$ and $t_i'$ are respectively equal. At the same time, they have the same eigenvalues. This point is also the topological phase transition point of $H_d^1$, so the transition point is not influenced by the other diagonal terms of the NPR Hamiltonian. Also, $H_d$ is a diagonal blocking matrix, and each block energy spectrum doesn't interfere with each other. In conclusion, it has the same topological phase as $H_d^1$.

For example, when $\lambda = 2$, the system is the SSH model. The form of $H_{d,SSH}^i$ is as follows,

$$H_{d,SSH}^i = \begin{pmatrix} 0 & \left(\frac{t_2}{t_1}\right)^{\frac{i-1}{N}} \cdot t_1 + \left(\frac{t_1}{t_2}\right)^{\frac{i-1}{N}} \cdot t_2 e^{-ik} \\ \left(\frac{t_2}{t_1}\right)^{\frac{i-1}{N}} \cdot t_1 + \left(\frac{t_1}{t_2}\right)^{\frac{i-1}{N}} \cdot t_2 e^{ik} & 0 \end{pmatrix}. \tag{A16}$$

According to the winding number function,

$$w := \frac{1}{2\pi i} \oint_{BZ} dk \frac{d}{dk} \log h(k), \tag{A17}$$

where $h(k) = \left(\frac{t_2}{t_1}\right)^{\frac{i-1}{N}} \cdot t_1 + \left(\frac{t_1}{t_2}\right)^{\frac{i-1}{N}} \cdot t_2 e^{-ik}$. According to the residue theorem, the winding number is 1 in the case of $\left(\frac{t_2}{t_1}\right)^{\frac{i-1}{N}} \cdot t_1 < \left(\frac{t_1}{t_2}\right)^{\frac{i-1}{N}} \cdot t_2$, and $t_2 = t_1$ is the singularity. If $N$ is big enough, there will be a value of $i$ making winding number $\omega = 1$ when $t_2 > t_1$, and then $H_{d,SSH}^i$ ($i \neq 1$) have the same topological phase as $H_{d,SSH}^1$. And for the other blocks, because the singularity $t_2 = t_1$ still exists, they will not influence the topological point of $H_d$. There will be more topological blacks. This also explains why there is a topological enhancement effect.

## APPENDIX B: APPLICATION OF NPR OPERATION

We use the SSH model to propose and prove the applicability of the NPR operation in the study of the topological phase. Next, we will use the Kagome lattice [29] model and the Kitaev model [42] to verify the application of the NPR operation.

(1) Kagome model

The Hamiltonian of Kagome lattice model is as follows,

$$H_{Kagome} = \begin{pmatrix} 0 & t_1 + t_2 e^{-i(k_x/2+\sqrt{3}k_y/2)} & t_1 + t_2 e^{-ik_x} \\ t_1 + t_2 e^{i(k_x/2+\sqrt{3}k_y/2)} & 0 & t_1 + t_2 e^{i(-k_x/2+\sqrt{3}k_y/2)} \\ t_1 + t_2 e^{ik_x} & t_1 + t_2 e^{i(k_x/2-\sqrt{3}k_y/2)} & 0 \end{pmatrix}, \quad (B1)$$

where $t$ is the coupling parameter. According to the NPR operation, the NPR Kagome model is as follows,

$$\sqrt[N]{H_{Kagome}} = \sum_{i=1}^{3N-1} t_1 \cdot a_i^\dagger a_{i+1} + t_1 \cdot a_1^\dagger a_{3N} + t_2 \cdot a_1^\dagger a_{3N+1}$$

$$+ \sum_{j=1}^{3} t_2 \cdot a_{(j-1)N+1}^\dagger a_{3N+(j-1)(N-1)+1}$$

$$+ \sum_{j=1}^{3} \sum_{i=1}^{N-1} t_2 \cdot a_{3N+(j-1)(N-1)+i}^\dagger a_{3N+(j-1)(N-1)+i+1}$$

$$+ t_2 \cdot e^{i(k_x/2+\sqrt{3}k_y/2)} a_{N+1}^\dagger a_{4N-1}$$

$$+ t_2 \cdot e^{i(k_x/2-\sqrt{3}k_y/2)} a_{2N+1}^\dagger a_{5N-2}$$

$$+ t_2 \cdot e^{-ik_x} a_1^\dagger a_{6N-3}. \quad (B2)$$

For example, $N = 3$, we can calculate the eigenvalues of the Hamiltonian, and give the energy spectrum. In this way, the spectrum of the original model is provided as shown in Fig. B1(a) from which we can find the topological region is $-1 < \frac{t_1}{t_2} < 0.5$. In the same parameters calculation of 3rd root model shown in Fig. B1(b), we can see the topological region is the same as the original one, $-1 < \frac{t_1}{t_2} < 0.5$.

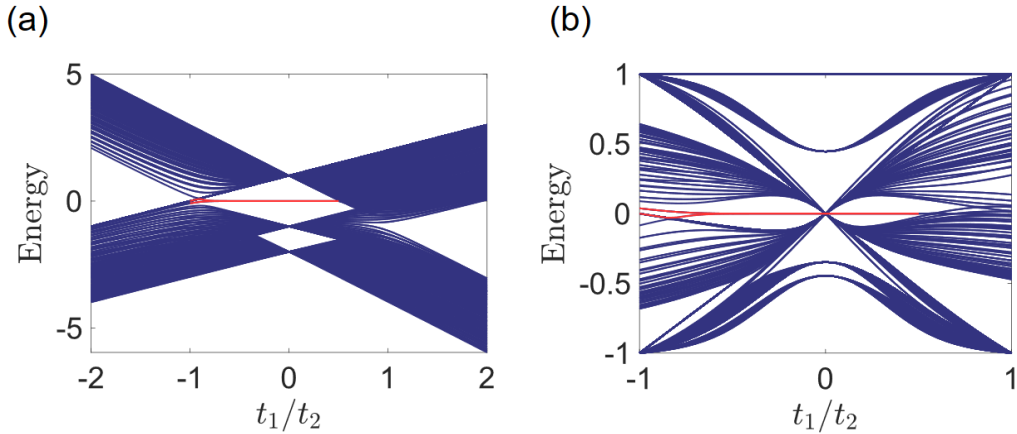

Fig. B1. Topological phase characterization of the NPR Kagome lattice model. An open triangle layout with size n = 50 (unit cell). $t_2 = 1$. (a) The energy spectrum of the Kagome lattice model. The topological region is marked in red, $-1 <$

$\frac{t_1}{t_2} < 0.5$. (b) The energy spectrum of 3$^{rd}$ root Kagome lattice model. The topological edge states are marked in red and the region is $-1 < \frac{t_1}{t_2} < 0.5$.

(2) Kitaev model

The Hamiltonian of Kitaev model is as follows,

$$H_{Kitaev} = (2t\cos k - \mu)\sigma_z + 2\Delta\sin k\sigma_x, \tag{B3}$$

where $t$, $\mu$ and $\Delta$ are couplings between different lattices. $\sigma$ is the Pauli operator. According to the NPR operation, the NPR Kitaev model is as follows,

$$H_{Kitaev} = t_1\left(a_1^\dagger a_3 + a_{N+1}^\dagger a_1 e^{ik}\right) + \Delta_1 a_1^\dagger a_{N+2} + \Delta_2 a_{N+3}^\dagger a_1 e^{ik}$$

$$+ t_2 a_2^\dagger a_{3N} + t_2 a_{4N-2}^\dagger a_2 e^{ik} + \Delta_2 a_2^\dagger a_{3N-1} + \Delta_2 a_{3N-2}^\dagger a_2 e^{ik}$$

$$+ \sum_{i=1}^{N-2}(t_1 a_{2+i}^\dagger a_{3+i} + t_2 a_{3N-1+i}^\dagger a_{3N+i})$$

$$+ \sum_{i=1}^{N-2}(\Delta_1 a_{N+2i}^\dagger a_{N+2+2i} + \Delta_2 a_{N+3+2i}^\dagger a_{N+3}), \tag{B4}$$

where $t_1 = \sqrt[N]{t}, t_2 = \sqrt[N]{-t}, \Delta_1 = e^{i\pi/2N}\sqrt[N]{\Delta}, \Delta_2 = e^{-i\pi/2N}\sqrt[N]{\Delta}$. And we can calculate the eigen values of the Hamiltonian, and give the energy spectrum. In this way, the spectrum of the original model is given as shown in Fig. B2(a) from which we can find the topological energy band marked in red. In the same parameters calculation of 3th root model shown in Fig. B2(b), we can see the topological region is same as the original one.

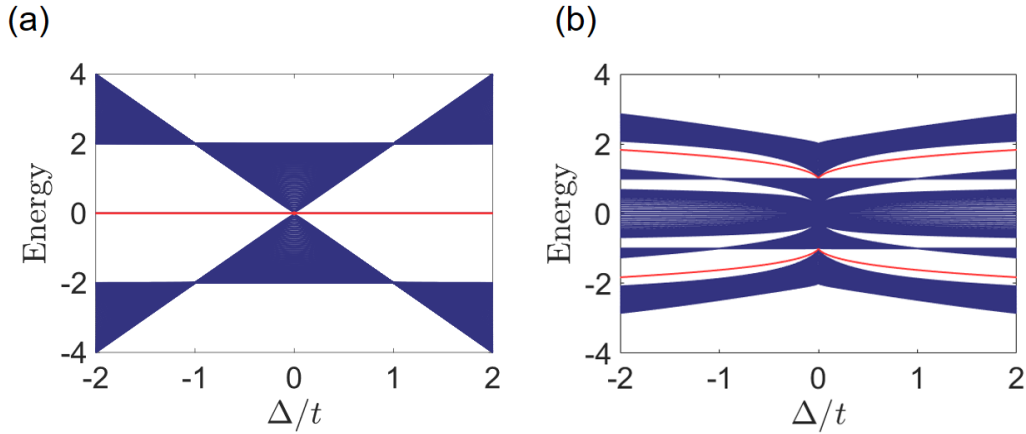

Fig. B2. Topological phase characterization of the Kitaev model (a) The energy spectrum of Kitaev model. The zero mode is marked in red. (b) The energy spectrum of the NPR Kitaev model. Same as (a), the topological mode is marked in red. But it's in finite energy.

**REFERENCES**

[1] M. Z. Hasan and C. L. Kane, Topological insulators, Rev. Mod. Phys. 82, 3045 (2010).

[2] X.-L. Qi and S.-C. Zhang, Topological insulators and superconductors, Rev. Mod. Phys. 83, 1057 (2011).

[3] W. A. Benalcazar, B. A. Bernevig, and T. L. Hughes, Quantized electric multipole insulators, Science 357, 61 (2017).

[4] H. Shen, B. Zhen and L. Fu, Topological Band Theory for Non-Hermitian Hamiltonians, Phys. Rev. Lett. 120,146402 (2018).

[5] J. K. Asbóth, L. Oroszlány, A. Pályi. The Su-Schrieffer-Heeger (SSH) model. In: A Short Course on Topological Insulators. Lecture Notes in Physics, vol 919. Springer, Cham. (2016)

[6] B. Midya, W. Walasik, N. M. Litchinitser, and L. Feng, Supercharge optical arrays, Opt. Lett. 43, 4927 (2018).

[7] G. Pelegrí, A. M. Marques, V. Ahufinger, J. Mompart, and R. G. Dias, Second-order topological corner states with ultracold atoms carrying orbital angular momentum in optical lattices, Phys. Rev. B 100, 205109 (2019).

[8] G. Pelegrí, A. M. Marques, R. G. Dias, A. J. Daley, V. Ahufinger, and J. Mompart, Topological edge states with ultracold atoms carrying orbital angular momentum in a diamond chain, Phys. Rev. A 99, 023612 (2019).

[9] C. Poli, M. Bellec, U. Kuhl, F. Mortessagne and H. Schomerus, Selective enhancement of topologically induced interface states in a dielectric resonator chain, Nat. Commun. 6, 6710 (2015).

[10] T. E. Lee, Anomalous Edge State in a Non-Hermitian Lattice, Phys. Rev. Lett. 116, 133903 (2016).

[11] S. Weimann, M. Kremer, Y. Plotnik, Y. Lumer, S. Nolte, K. G. Makris, M. Segev, M. C. Rechtsman, A. Szameit, Topologically protected bound states in photonic parity–time-symmetric crystals, Nat. Mat. 16, 433 (2017).

[12] H. Zhao, P. Miao, M. H. Teimourpour, S. Malzard, R. El-Ganainy, H. Schomerus, L. Feng, Topological hybrid silicon microlasers, Nat. Commun. 9, 981 (2018).

[13] G. Harari, M. A. Bandres, Y. Lumer, M. C. Rechtsman, Y. D. Chong, M. Khajavikhan, D. N. Christodoulides, M. Segev, Science 359(6381), Topological insulator laser: theory, eaar4003 (2018).

[14] M. A. Bandres, S. Wittek, G. Harari, M. Parto, J. Ren, M. Segev, D. N. Christodoulides, M. Khajavikhan, Science 359(6381), Topological insulator laser: Experiments, eaar4005 (2018).

[15] D. Leykam, K. Y. Bliokh, C. Huang, Y. D. Chong, F. Nori, Edge modes, degeneracies, and topological numbers in non-Hermitian systems, Phys. Rev. Lett. 118, 040401 (2017).

[16] S. Lieu, Topological phases in the non-Hermitian Su-Schrieffer-Heeger model, Phys. Rev. B, 97, 045106 (2018).

[17] M.-A. Miri and A. Alù, Exceptional points in optics and photonics, Science 363, 7709 (2019).

[18] Okuma, N., et al. Topological Origin of Non-Hermitian Skin Effects, Phys. Rev. Lett. 124, 086801 (2020).

[19] N. Okuma, K. Kawabata, K. Shiozaki, and M. Sato, Topological Origin of Non-Hermitian Skin Effects, Phys. Rev. Lett. 124,086801 (2020).

[20] L. Li, C. H. Lee, and J. Gong, Topological Switch for Non-Hermitian Skin Effect in Cold-Atom Systems with Loss, Phys. Rev. Lett. 124, 250402 (2020).

[21] Z. Xiao, H. Li, T. Kottos, A. Alù, Enhanced sensing and nondegraded thermal noise performance based on P T-symmetric electronic circuits with a sixth-order exceptional point. Physical Review Letters, 123, 213901 (2019).

[22] J. C. Budich and E. J. Bergholtz, Non-Hermitian topological sensors. Physical Review Letters, 125, 180403 (2020).

[23] S. Weidemann, et al, Topological funneling of light, Science 368, 311-314 (2020).

[24] C. H. Lee, L. Li, and J. Gong, Hybrid Higher-Order Skin-Topological Modes in Nonreciprocal Systems, Phys. Rev. Lett. 123, 016805 (2019).

[25] X. W. Luo and C. Zhang, Higher-Order Topological Corner States Induced by Gain and Loss, Phys. Rev. Lett. 123, 073601 (2019).

[26] T. Liu, Y. R. Zhang, Q. Ai, Z. Gong, K. Kawabata, M. Ueda and F. Nori, Second-Order Topological Phases in Non-Hermitian Systems, Phys. Rev. Lett. 122, 076801 (2019).

[27] C. W. Peterson, W. A. Benalcazar, T. L. Hughes and G. Bahl, A quantized microwave quadrupole insulator with topologically protected corner states, Nature 555, 346 (2018).

[28] M. Serra-Garcia, V. Peri, R. Süsstrunk, O. R. Bilal, T. Larsen, L. G. Villanueva and S. D. Huber, Observation of a phononic quadrupole topological insulator, Nature 555, 342 (2018).

[29] H. Xue, Y. Yang, F. Gao, Y. Chong and B. Zhang, Acoustic higher-order topological insulator on a kagome lattice. Nat. Mater, 18, 108–112 (2019).

[30] S. Mittal, V.V. Orre, G. Zhu, M. A. Gorlach, A. Poddubny and M. Hafezi, Photonic quadrupole topological phases, Nat. photonics 13, 692-696 (2019).

[31] W. Zhang, X. Xie, H. Hao, J. Dang, S. Xiao, S. Shi, H. Ni, Z. Niu, C. Wang, K. Jin, X. Zhang and X. Xu, Low-threshold topological nanolasers based on the second-order corner state, Light: Science & Applications, 9 (1), 1-6 (2020).

[32] R. El-Ganainy, K. G. Makris, M. Khajavikhan, Z. H. Musslimani, S. Rotter, D. N. Christodoulides, Non-Hermitian physics and PT symmetry, Nat. Phys. 14. 11 (2018).

[33] H. Shen, B. Zhen, L. Fu, Topological band theory for non-Hermitian Hamiltonians, Phys. Rev. Lett. 120, 146402 (2018).

[34] X. Ni, M. Weiner, A. Alu and A. B. Khanikaev, Observation of higher-order topological acoustic states protected by generalized chiral symmetry, Nat. Mater. 18, 113–120 (2019).

[35] A. M. Marques and R. G. Dias, One-dimensional topological insulators with noncentered inversion symmetry axis, Phys. Rev. B 100, 041104 (2019).

[36] L. Jin, Z. Song, Bulk-boundary correspondence in a non-Hermitian system in one dimension with chiral inversion symmetry, Phys. Rev. B 99, 081103 (2019).

[37] N. Moiseyev, Non-Hermitian Quantum Mechanics, Cambridge University Press (2011).

[38] V. M. Martinez Alvarez, J. E. Barrios Vargas and L. E. F. Foa Torres, Non-Hermitian robust edge states in one dimension: Anomalous localization and eigenspace condensation at exceptional points, Phys. Rev. B. 97, 121401 (2018).

[39] C. Wang, A. C. Potter and T. Senthil, Classification of Interacting Electronic Topological Insulators in Three Dimensions, Science 343, 629 (2014).

[40] J. Maciejko and G. A. Fiete, Fractionalized topological insulators, Nat. Phys. 11, 385 (2015).

[41] J. Arkinstall, M. H. Teimourpour, L. Feng, R. El-Ganainy, and H. Schomerus, Topological tight-binding models from nontrivial square roots, Phys. Rev. B 95, 165109 (2017).

[42] M. Ezawa, Systematic construction of square-root topological insulators and superconductors, Phys. Rev. Res. 2, 033397 (2020).

[43] T. Mizoguchi, Y. Kuno, and Y. Hatsugai, Square-root higherorder topological insulator on a decorated honeycomb lattice Phys. Rev. A 102, 033527 (2020).

[44] M. Kremer, I. Petrides, E. Meyer, M. Heinrich, O. Zilberberg, and A. Szameit, A square-root topological insulator with non-quantized indices realized with photonic Aharonov-Bohm cages, Nat. Commun. 11, 907 (2020).

[45] Z. Zhang, M. H. Teimourpour, J. Arkinstall, M. Pan, P. Miao, H. Schomerus, R. El-Ganainy, and L. Feng, Experimental realization of multiple topological edge states in a 1D photonic lattice, Laser Photonics Rev. 13, 1800202 (2019).

[46] L. Song, H. Yang, Y. Cao, and P. Yan, Realization of the square root higher-order topological insulator in electric circuits, Nano Lett. 20, 7566 (2020)

[47] S. Yao and Z. Wang, Edge states and topological invariants of non-Hermitian systems, Phys. Rev. Lett. 121, 086803 (2018).

[48] S. Yao and F. Song, and Z. Wang, Non-Hermitian Chern Bands, Phys.Rev.Lett.2018,121(136802).

[49] Z. Gong, Y. Ashida, K. Kawabata, K. Takasan, S. Higashikawa and M. Ueda, Topological phases of non-Hermitian systems. Phys. Rev. X 8, 031079 (2018).

[50] S. Imhof, C. Berger, F. Bayer, J. Brehm, L. W. Molenkamp, T. Kiessling, F. Schindler, C. H. Lee, M. Greiter, T. Neupert and R. Thomale, Topolectrical-circuit realization of topological corner modes, Nat. Phys. 14, 925 (2018).

[51] C. H. Lee, S. Imhof, C. Berger, F. Bayer, J. Brehm, L. W. Molenkamp, T. Kiessling and R. Thomale, Topolectrical circuits. Communications Physics, 1, 39 (2018).

[52] V. Albert, L. I. Glazman and L. Jiang, Topological properties of linear circuit lattices, Phys. Rev. Lett. 114, 173902 (2015).

[53] M. Ezawa, Electric circuit simulations of nth-Chern-number insulators in 2n-dimensional space and their non-Hermitian generalizations for arbitrary, Phys. Rev. B 100, 075423 (2019).

[54] T. Hofmann, T. Helbig, C. H. Lee, M. Greiter and R. Thomale, Chiral voltage propagation and calibration in a topolectrical Chern circuit. Phys. Rev. Lett. 122, 247702 (2019).

[55] X. X. Zhang and M. Franz, Non-Hermitian Exceptional Landau Quantization in Electric Circuit. Phys. Rev. Lett. 124, 046401 (2020).

[56] R. Yu, Y. Zhao and A. P. Schnuder, 4D spinless topological insulator in a periodic electric circuit. National Science Review, nwaa065 (2020).

[57] J. Bao, D. Zou, W. Zhang, W. He, H. Sun and X. Zhang, Topolectrical-circuit octupole insulator with topologically protected corner states. Phys. Rev. B 100, 201406 (2019).

[58] N. Olekhno, E. I. Kretov, A. A. Stepanenko, P. A. Ivanova, V. V. Yaroshenko, E. M. Puhtina, D. S. Filonov, B. Cappello, L. Matekovits and M. A. Gorlach, Topological edge states of interacting photon pairs realized in a topolectrical circuit. Nat. Commun. 11, 1436 (2020).

[59] Y. Wang, H. M. Price, B. Zhang, Y. D. Chong, Circuit Realization of a Four-Dimensional Topological Insulator. Nat. Commun. 11, 2356 (2020).

[60] J. Ningyuan, C. Owens, A. Sommer, D. Schuster, J. Simon, Time and site resolved dynamics in a topological circuit, Phys. Rev. X 5, 021031 (2015).

[61] W. Zhang, D. Zou, J. Bao, W. He, Q. Pei, H. Sun, X. Zhang, Topolectrical-circuit realization of a four-dimensional hexadecapole insulator, Phys. Rev. B 102, 100102 (2020).

[62] C. H. Lee, A. Sutrisno, T. Hofmann, T. Helbig, Y. Liu, Y. Ang, L. Ang, X. Zhang, M. Greiter and R. Thomale, Imaging nodal knots in momentum space through topolectrical circuits, Nature communications 11, 1-13 (2020).

[63] Z. Yang, K. Zhang, C. Fang and J. Hu, Non-Hermitian Bulk-Boundary Correspondence and Auxiliary Generalized Brillouin Zone Theory, Phys. Rev. Lett. 125, 226402 (2020).

[64] K. Zhang, Z. Yang, and C. Fang, Correspondence between Winding Numbers and Skin Modes in Non-Hermitian Systems, Phys. Rev. Lett. 125, 126402 (2020).

[65] K. Kawabata, K. Shiozaki, M. Ueda, and M. Sato, Symmetry and topology in non-Hermitian physics, Phys.Rev. X 9, 041015 (2019).

[66] L. Xiao, T. Deng, K. Wang, G. Zhu, Z. Wang, W. Yi and P. Xue, Observation of non-Hermitian bulk-boundary correspondence in quantum dynamics, Nat. Phys. 16, 761 (2020).

[67] A. Ghatak, A., et al. Observation of non-Hermitian topology and its bulk-edge correspondence in an active mechanical metamaterial, Proc. Natl. Acad. Sci. USA 117, 29561 (2020).

[68] T. Helbig, T. Hofmann, S. Imhof, M. Abdelghany, T. Kiessling, L. W. Molenkamp, C. H. Lee, A. Szameit, M. Greiter and R. Thomale, Generalized bulk–boundary correspondence in non-Hermitian topolectrical circuits, Nat. Phys. 16, 747 (2020).

[69] L. Zhou, R. W. Bomantara, and S. Wu, qth-root non-Hermitian Floquet topological insulators, arXiv:2203.09838v1.

Wenyuan Deng[*], Tian Chen[*], and Xiangdong Zhang[+]

[1]Key Laboratory of advanced optoelectronic quantum architecture and measurements of Ministry of Education, Beijing Key Laboratory of Nanophotonics & Ultrafine Optoelectronic Systems, School of Physics, Beijing Institute of Technology, 100081, Beijing, China

[*]These authors contributed equally to this work. [+*]Author to whom any correspondence should be addressed. E-mail: zhangxd@bit.edu.cn; chentian@bit.edu.cn

## S1 THE CALCULATION OF THE WINDING NUMBER

In Hermitian and non-Hermitian cases, we can obtain the topological phase boundary by carrying out a detailed calculation of the winding number. The winding number of the NPR Hermitian SSH model calculated by Eq. (A17) is shown in Fig. S1(a). We can find that the boundary is still $t_1 = t_2$. In non-Hermitian cases, we can calculate the winding number of the SSH model by

$$w := \frac{1}{2\pi}\oint_C \frac{d}{dx} arg[H(x) - E_b] \, dx. \tag{S5}$$

As the calculation result of the NPR non-Hermitian SSH model in Fig. 1S(b), we can see that the region of the NPR model's topological region is the still same as the original model, $-\sqrt{t_2^2 + \gamma^2} < t_1 < \sqrt{t_2^2 + \gamma^2}$

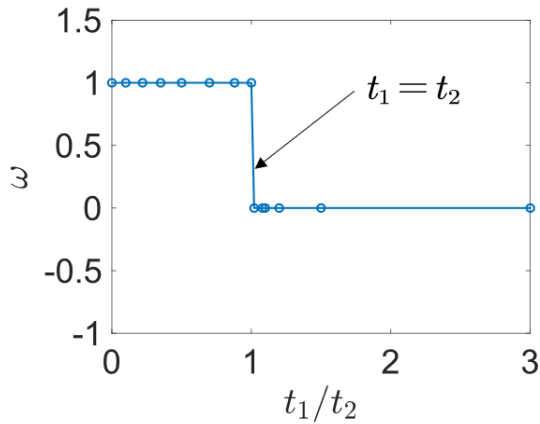
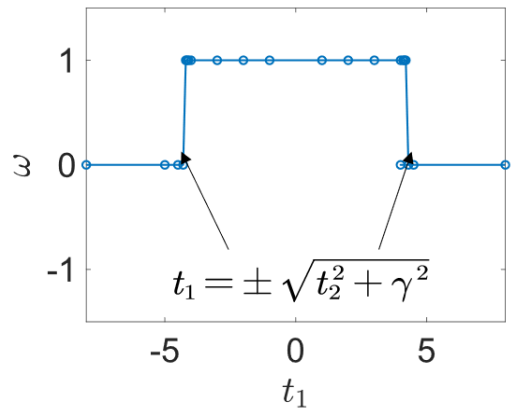

Fig. S1. The numerical calculation results of the NPR SSH model. (a) The winding number of the Hermitian cases. $N = 3, t_2 = 1$. The winding number is 1 when $t_1 < t_2$. (b) The winding number of the non-Hermitian cases. $N = 3, t_2 = 4, \gamma = \frac{4}{3}$. The region for the non-trivial topological phase is $-\sqrt{t_2^2 + \gamma^2} < t_1 < \sqrt{t_2^2 + \gamma^2}$

## S2 THE DESIGN OF CIRCUIT EXPERIMENT

In the circuit experiment, the variation of current and voltage follows Kirchhoff's law. Therefore, for the RLC circuit composed of resistors, inductors, capacitors and other components. The changed behavior of current and voltage is controlled by its circuit Laplace operator, which is similar to the effect of Hamiltonian describing the energy of the physical system.

Any circuit network can be represented by a graph whose nodes and edges correspond to circuit nodes and connecting lines or components. The variation of the circuit can be determined by Kirchhoff's current law as

$$I_A = \sum_i c_{Ai}(V_A - V_i) + w_A V_A, \tag{S6}$$

where $I_A$ and $V_A$ are respectively the current flowing into the node A and the potential of the node A, and i is the index of the other nodes connected. $c_{Ai}$ is the admittance between node A and other nodes, $w_A$ is the admittance between node A and ground. According to the conservation of current, it equals to the total current from the node to all other nodes connected by non-zero conductivity, plus the current flowing into the ground through the path with impedance. We can use the above theory to construct the topological circuit to show the boundary states in experiments.

### A. Hermitian Circuit

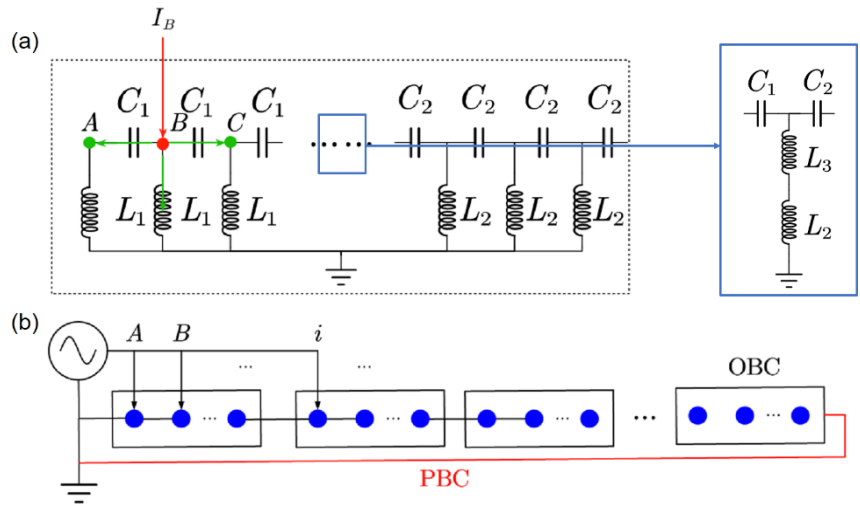

Fig. S2. Circuit diagram. (a) The Hermitian circuit. A current feed is imposed on a measurement node B marked in red. The current flow directions are indicated by green arrows. A particular measurement node is marked in blue, and this design

is for eliminating diagonal terms in the admittance matrix. (b) The method of measuring experiments. We set the current feed on every measurement node. And every time the feed is set, we measure the trunk current and the voltage of each node. With the coupling line marked in red, it's an experiment in the PBC. Otherwise, it's an experiment in the OBC.

In Fig. S4(a), we give the design of the Hermitian circuit. As the measuring node B, according to Eq. (S2), the inflow current marked in red is equal to the sum of outflow currents marked in green. The Kirchhoff's current law is given by Eq. (S2) as

$$I_B = (V_B - V_C) \cdot i\omega C_1 + (V_B - V_A) \cdot i\omega C_1 + V_B \cdot \frac{1}{i\omega L_1}, \tag{S7}$$

where B and C indicate the nodes connected with node A. $i\omega C_1$ and $\frac{1}{i\omega L_1}$ are the admittance of different connections. Also, for nodes A and C, the Kirchhoff's current functions are

$$0 = (V_A - V_B) \cdot i\omega C_1 + V_A \cdot \frac{1}{i\omega L_1}, \tag{S8}$$

$$0 = (V_B - V_C) \cdot i\omega C_1 + (V_B - V_A) \cdot i\omega C_1 + V_B \cdot \frac{1}{i\omega L_1}. \tag{S9}$$

If we have all the nodes' Kirchhoff's current functions, we can give the admittance matrix of the whole circuit as

$$J \begin{bmatrix} V_A \\ V_B \\ \vdots \\ V_i \\ \vdots \end{bmatrix} = \begin{bmatrix} 0 \\ I_B \\ \vdots \\ 0 \\ \vdots \end{bmatrix}, \tag{S10}$$

the admittance matrix is as

$$J = \begin{pmatrix} \ddots & & & & & & & \\ 2\omega C_1 - \frac{1}{\omega L_1} & -\omega C_1 & & & & & & \\ -\omega C_1 & 2\omega C_1 - \frac{1}{i\omega L_1} & -\omega C_1 & & & & & \\ & -\omega C_1 & 2\omega C_1 - \frac{1}{\omega L_1} & -\omega C_1 & & & & \\ & & -\omega C_1 & \omega(C_1+C_2) - \frac{1}{\omega(L_2+L_3)} & -\omega C_2 & & & \\ & & & -\omega C_2 & 2\omega C_2 - \omega L_2 & -\omega C_2 & & \\ & & & & -\omega C_2 & 2\omega C_2 - \omega L_2 & \omega C_2 & \\ & & & & & \omega C_2 & 2\omega C_2 - \omega L_2 & \\ & & & \ddots & & & & \end{pmatrix}, \tag{S11}$$

where $\omega(C_1+C_2) - \frac{1}{\omega(L_2+L_3)}$ is a particular node, the term connecting ground is indicated by the blue square in Fig. S4(a). As Eq. (S7), to match it with the NPR quantum system introduced before, we need to take the diagonal terms to zero. The relationship between every component is given as

$$\begin{cases} 2\omega C_1 = \frac{1}{\omega L_1} \\ \omega(C_1+C_2) = \frac{1}{\omega(L_2+L_3)} \\ \cdots \end{cases} \tag{S12}$$

It is the basis for selecting the experimental frequency range. This circuit designed above can show some properties of the topological phase. To measure the boundary states in the Hermitian circuit, we design a series of measurement experiments, as shown in Fig. S2(b). We also change the feed node to measure the boundary states in the circuit. After measurement, the admittance matrix can be calculated by

$$J[\vec{V}_1 \quad \vec{V}_2 \quad \cdots \quad \vec{V}_n] = \begin{bmatrix} I_A & & & \\ & I_B & & \\ & & \ddots & \\ & & & I_n \end{bmatrix}, \quad (S13)$$

where $\vec{V}$ is the column vector of the voltages in one measuring experiment when the feed is not changed. $I$ is the trunk current also in one measuring experiment. With the coupling line marked in red, we can execute the experiment in the PBC, and we can have the complex spectrum of the system. Without the coupling line marked in red, we can execute the experiment in the OBC, and the density of states distribution can be obtained.

## B. Non-Hermitian Circuit

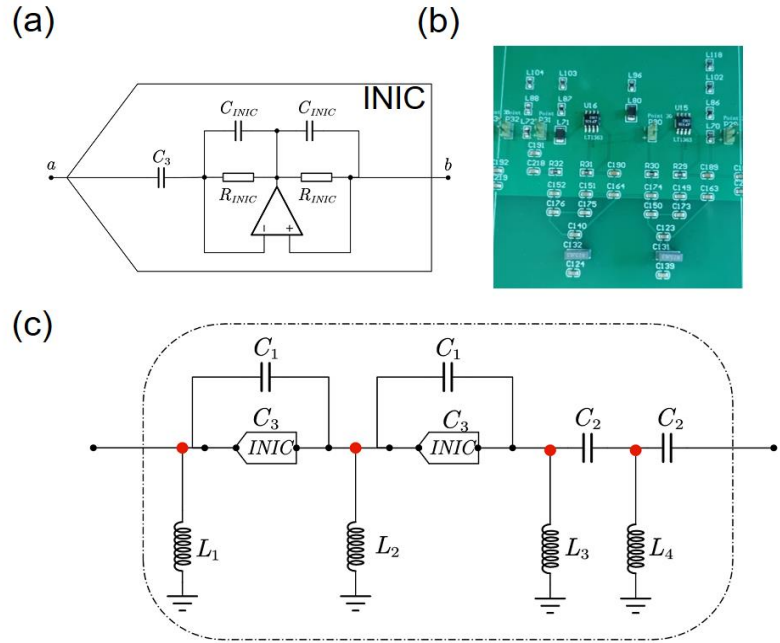

Fig.S3. The experiment design of the NPR non-Hermitian SSH model when $N = 2$. (a) Diagram of negative impedance circuit. The structure is a negative impedance circuit. The feedback characteristics of the operational amplifier are used, resulting in a phase difference of $\pi$ between the impedance from a to b and from b to a. (b) Circuit board cutout of one unit-cell of the square root model. (c) The diagram of the square root non-Hermitian circuit. Here, $C_1 = 197nF, C_2 = 200nF, C_3 = 34nF, L_1 = 65.3\mu H, L_2 = 71.4\mu H, L_3 = 77.5\mu H, L_4 = 70.4\mu H$ and the negative impedance circuit $C_{INIC} = 200nF, R_{INIC} = 200\Omega$.

Due to the dissipation in the non-Hermitian system, the energy transfer between some nodes is not equal, so we need to introduce the INIC into the circuit system. As shown in Fig. S3(a), the admittance between nodes a and b is opposite to each other, and the phase difference makes the admittance directional, so that the circuit system corresponding to the non-Hermitian model can be realized, as shown in Fig. S3(b). We use its directional characteristics to simulate the unbalanced nature of energy transfer in the non-Hermitian circuits. Because of the use of the operational amplifier, the circuit oscillation appears. To eliminate part of the oscillation and reduce experimental errors, some capacitors are added in experiments that don't affect the admittance matrix properties of the system.

As shown in Fig. S3(c), the square root non-Hermitian circuit, same as Eq. (S6), we can give the admittance matrix as

$$J' = \begin{pmatrix} & \cdots & & & \\ & \omega(C_1-C_3) - \frac{1}{i\omega L_1} & -\omega(C_1-C_3) & & \\ \vdots & -\omega(C_1+C_3) & 2\omega C_1 - \frac{1}{\omega L_2} & -\omega(C_1-C_3) & \vdots \\ & -\omega(C_1+C_3) & \omega(C_1+C_2+C_3) - \frac{1}{\omega L_3} & -\omega C_2 & \\ & & -\omega C_2 & 2\omega C_2 - \frac{1}{\omega L_4} & \\ & & \cdots & & \end{pmatrix}. \quad (S14)$$

Then we can also execute the experiment as in the Hermitian case shown in Fig. S5(b). In this way, we can get the voltage distribution with the skin effect and get the admittance matrix.

## S3 THE CORRECTION OF THE NPR OPERATION IN THE PBC

Because of the unusual BBC of non-Hermitian cases, we need to verify the same properties between original and NPR models not only in the OBC but also in the PBC. In this process, we find that the spectra of NPR models in the momentum space drift away from the original position. For example, Fig. S4(a), shows the complex spectrum of the 4$^{th}$ root SSH model has a gap for the case of $t_1 = 5$. But the transition point of the SSH model is at $t_1 = \frac{16}{3} [t_1 = t_2 + \gamma ]$. Compared with the case of $t_1 = 4$ shown in Fig. S4(b), the transition points are still in the region of [4,5], which has drifted after the NPR operation. To make the transition points back to the original value in the numerical calculation, we designed a correction scheme adding unequal site energy for non-Hermitian SSH model as

$$H_r^c = H_r' + H_\varepsilon, \quad (S15)$$

where $H_\varepsilon$ represent $\varepsilon(a_1^\dagger a_1 + a_{N+1}^\dagger a_{N+1})$ and $\varepsilon = 2(N-2)i\sqrt{\sqrt[N]{t_1 t_1'} + \sqrt[N]{t_2^2}}(N \geqslant 2)$. According to Eq. (S11), we give the complex spectrum of $N = 4$ in the PBC shown in Fig. S4(c) and (d). By judging whether the energy ring is disconnected, we can get the transition points of $N = 4$ which are the same as the original model [$t_2 - \gamma < t_1 < t_2 + \gamma$ and $-t_2 - \gamma < t_1 < -t_2 + \gamma$].

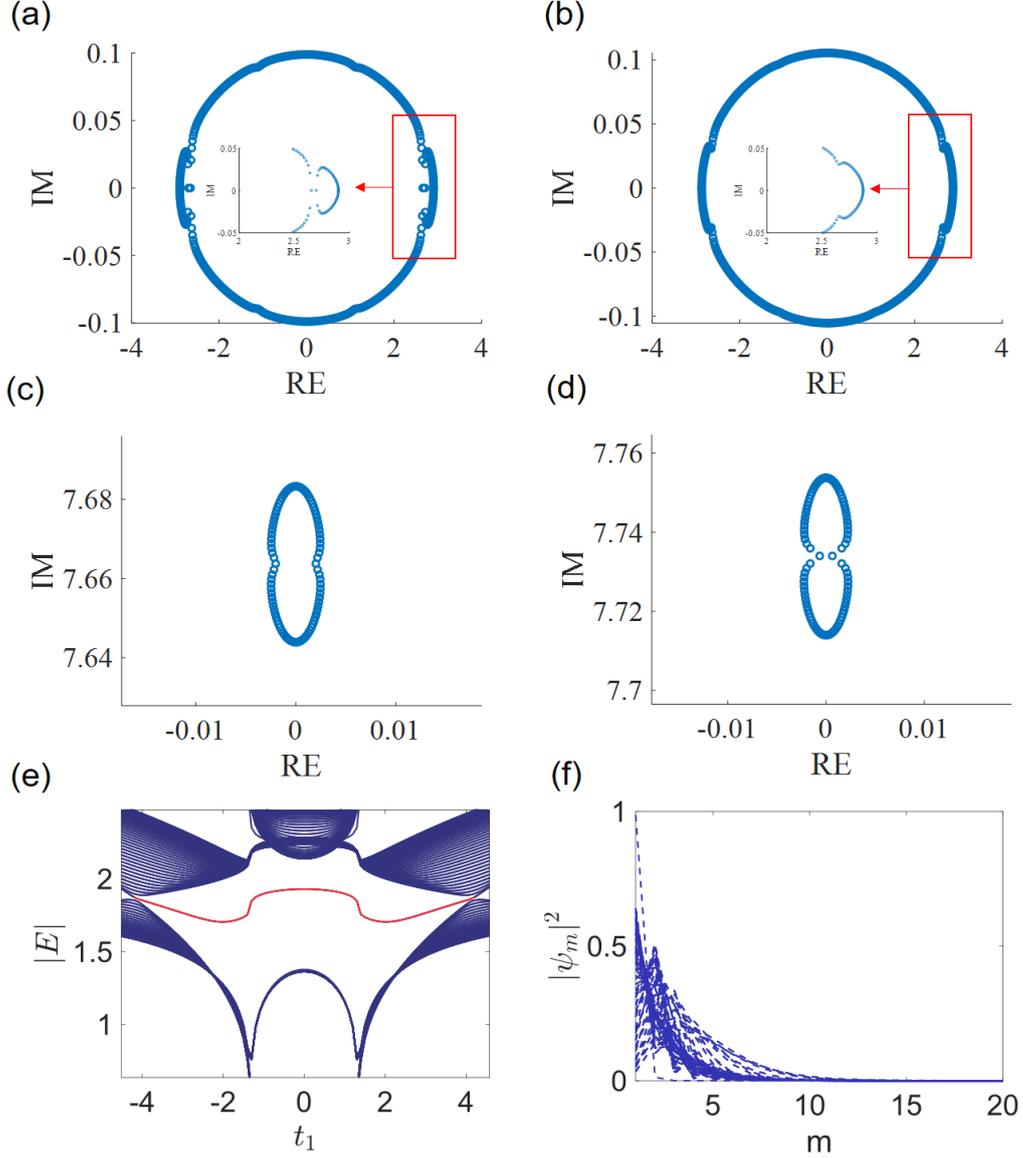

Fig.S4. Comparison of corrected and uncorrected energy spectra. (a) Energy spectra without correction when $N = 4, t_1 = 5, t_2 = 4, \gamma = 4/3$. The energy spectra indicate a non-topological phase. (b) The energy spectra without correction when $N = 4, t_1 = 4, t_2 = 4, \gamma = 4/3$. The energy spectra indicate a topological phase. (c) Energy spectra with correction when $N = 4, t_1 = 5, t_2 = 4, \gamma = 4/3$ indicating a topological phase. (d) Energy spectra with correction when $N = 4, t_1 = 16/3, t_2 = 4, \gamma = 4/3. t_1 = 16/3$. This corresponds to one of the critical topological regions.

After the correction, we can also calculate the DOS and spectrum in the OBC to ensure the properties of OBC are not changed by correction. For example, the Fig. S4(e) shows the spectrum of the 3$^{rd}$ root SSH model in the OBC, the transition points are still $t_1 \approx \pm\sqrt{t_2^2 + \gamma^2}$ that are the same as the original model. Corresponding to the spectrum in the OBC, we can also calculate the DOS of 3$^{rd}$ root SSH model (Fig. S4(f)). The skin effect can still be observed after correction.

## S4 THE ANALYSIS OF ENHANCEMENT ITEM $H_{enh}$

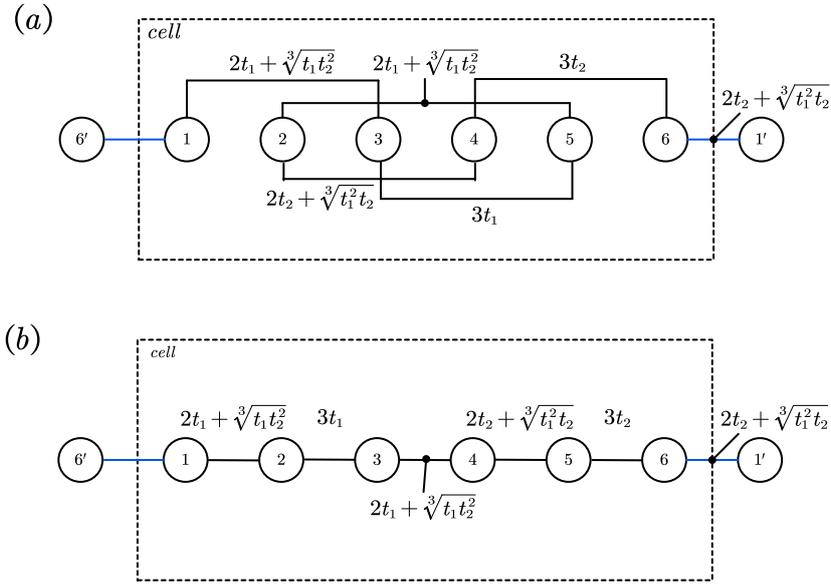

Fig. S5. The subdivided graph of the enhancement terms in the 3$^{rd}$ root model. Couplings between different unit cells are marked in blue. (a) The off-diagonal coupling terms are still in the form of a tight-binding model. The coupling strength is shown in the unit cell, which is indicated by a dotted line (b) The subdivided graph of (a) after unitary transformation. It's the same as the subdivided graph of diagonal terms.

For the enhancement terms, because the Nth power operation is matrix multiplication, the form of the off-diagonal term is the sum of the polynomial expansion of each block. We can analyze the enhancement term through the special case, $N = 3$. To compare the original model and enhancement terms of the NPR model, we give the Hamiltonian of the 3$^{rd}$ root model as

$$H_{hermit}^{N=3} = \begin{pmatrix} 0 & \sqrt[3]{t_1} & 0 & 0 & 0 & \sqrt[3]{t_2}e^{-ik} \\ \sqrt[3]{t_1} & 0 & \sqrt[3]{t_1} & 0 & 0 & 0 \\ 0 & \sqrt[3]{t_1} & 0 & \sqrt[3]{t_1} & 0 & 0 \\ 0 & 0 & \sqrt[3]{t_1} & 0 & \sqrt[3]{t_2} & 0 \\ 0 & 0 & 0 & \sqrt[3]{t_2} & 0 & \sqrt[3]{t_2} \\ \sqrt[3]{t_2}e^{ik} & 0 & 0 & 0 & \sqrt[3]{t_2} & 0 \end{pmatrix}, \quad (S16)$$

Using the unitary transformation $U_0 = \sum_{i=1}^{\lambda} \sum_{j=1}^{N} a^\dagger_{\lambda(j-1)+i} a_{N(i-1)+j}$ and Nth power operation in Eq. (S12), we can give its Nth power form as,

$$U_0 (H_{hermit}^{N=3})^3 U_0^{-1} = \begin{pmatrix} H'_\alpha H_\beta H_\beta + H_\alpha H_\alpha H'_\beta & H_\alpha H_\beta H_\alpha + H_\alpha H_\alpha H_\beta + H'_\alpha H'_\beta H_\alpha & H_\alpha H_\beta H'_\alpha + H'_\alpha H_\beta H_\alpha + H'_\alpha H'_\beta H'_\alpha \\ H_\alpha H_\beta H_\beta + H_\beta H_\alpha H_\beta + H_\beta H'_\alpha H'_\beta & H_\beta H'_\alpha H_\beta + H_\alpha H'_\beta H_\alpha & H_\alpha H_\beta H_\alpha + H_\beta H_\alpha H_\alpha + H'_\beta H'_\alpha H_\beta \\ H_\beta H_\alpha H'_\beta + H'_\beta H_\alpha H_\beta + H'_\beta H'_\alpha H'_\beta & H_\beta H_\beta H_\alpha + H_\beta H_\alpha H_\beta + H'_\beta H'_\alpha H_\beta & H_\beta H_\beta H'_\alpha + H'_\beta H_\alpha H_\alpha \end{pmatrix}, \quad (S17)$$

from which, we can give the form of $H_{enh}$. From the coupling graph of $H_{enh}$ as shown in Fig. S5(a), we can see it's still a tight binding form. To show its coupling distribution more clearly, we transform it from Fig. S5(a) to Fig. S5(b) by unitary transformation $U_0^{-1}$. The Hamiltonian of Fig. S5(b) is given as

$$U_0^{-1} H_{enh} U_0 = \begin{pmatrix} 0 & 2t_1 + \sqrt[3]{t_1 t_2^2} & 0 & 0 & 0 & \left(2t_2 + \sqrt[3]{t_1^2 t_2}\right) e^{-ik} \\ 2t_1 + \sqrt[3]{t_1 t_2^2} & 0 & 3t_1 & 0 & 0 & 0 \\ 0 & 3t_1 & 0 & 2t_1 + \sqrt[3]{t_1 t_2^2} & 0 & 0 \\ 0 & 0 & 2t_1 + \sqrt[3]{t_1 t_2^2} & 0 & 2t_2 + \sqrt[3]{t_1^2 t_2} & 0 \\ 0 & 0 & 0 & 2t_2 + \sqrt[3]{t_1^2 t_2} & 0 & 3t_2 \\ \left(2t_2 + \sqrt[3]{t_1^2 t_2}\right) e^{ik} & 0 & 0 & 0 & 3t_2 & 0 \end{pmatrix}, \quad (S18)$$

from which, we find that the coupling of $U_0^{-1} H_{enh} U_0$ is similar to the NPR model to a certain extent. We can understand that the enhancement term is the superposition of multiple identical models (including certain perturbations). The superposition of the diagonal term of the topological phase mentioned in Appendix A and the superposition of the above enhancement items finally lead to the enhancement of system localization.